# Critical Behavior and Anisotropy in Single Crystal $SrRuO_3$

*G. Jeffrey Snyder*
*Stanford University, Stanford CA USA*
*current address Northwestern University, Evanston IL USA*

**Abstract**
The Magnetization of Single Crystal $SrRuO_3$ is studied as a function of temperature along different crystallographic directions. The magnetocrystalline anisotropy and behavior near the critical transition temperature are analyzed in detail. The magnetization vs temperature is found to vary more like $T^2$ rather than $T^{3/2}$ expected for spin waves.

## Introduction

Strontium ruthenate, $SrRuO_3$, has many physical properties which make it unique among perovskite oxides. First, it is metallic in the undoped state [36, 126, 173-176]. This is even more unusual since the ruthenium in $SrRuO_3$ is in a high oxidation state; whereas many other metallic perovskites must be formed in reducing environments and therefore unstable in air at high temperatures. The remarkable chemical stability and simple chemical formula makes metallic $SrRuO_3$ quite attractive for use in epitaxial thin film heterostructures with other perovskite oxides when metallic layers are desired. Indeed, the nearly cubic $SrRuO_3$ [177] is often preferred over the more distorted $CaRuO_3$ when making structures such as electrodes for ferroelectrics or superconductor-normal metal-superconductor junctions.

The other striking feature of $SrRuO_3$ is its ferromagnetism with a reasonably high transition temperature (163 K) and large saturation moment ($> 1\mu_B$) [174, 175, 178, 179]. $SrRuO_3$ is the only ferromagnetic perovskite oxide of a 4*d* or 5*d* transition metal. Moreover, $SrRuO_3$ has the largest saturation moment known to arise from 4*d* electrons, making it more related to the iron group ferromagnetic metals (Fe, Co and Ni) than to the weak itinerant electron ferromagnets such as $ZrZn_2$. $SrRuO_3$ also has very strong cubic magnetic anisotropy, requiring magnetic fields in excess of 10 Tesla to saturate the magnetization in the hard directions. Such a strong anisotropy makes measuring even the simplest properties, such as saturation magnetization, difficult. It is the purpose of the present work to determine the magnetic properties of $SrRuO_3$ by measurements of magnetically-soft single crystals along the easy magnetic direction.

There have been several studies of $SrRuO_3$ in the past three decades, mostly on polycrystalline samples which are quite easy to prepare. $SrRuO_3$ has a very slightly distorted ($GdFeO_3$ type) perovskite structure. The deviation from perfect cubic perovskite is so small that it has often been undetected. The resistivity of polycrystalline, and epitaxial thin film $SrRuO_3$ shows a cusp in $d\rho/dT$ at the ferromagnetic Curie temperature $T_C$. This is commonly observed for metallic ferromagnets and is attributed to spin disorder scattering [125]. Reported $T_C$'s tend to vary from 150K to 165K.

The saturation magnetization of $SrRuO_3$ has been both difficult to measure and interpret. Low spin $Ru^{4+}$ in an octahedral coordination has four 4$d$ electrons in the $t_{2g}$ triply degenerate state, giving two paired and two unpaired electrons. Since the orbital component of angular momentum $J$ will be quenched, $J = S = 1$ is expected. The measured Curie constant of the paramagnetic state is consistent with this model (expected: $\mu_{eff} = \sqrt{2gJ(J+1)}\ \mu_B = 2.83\ \mu_B$; measured = 2.67 $\mu_B$ [175]). For a localized moment ferromagnet, the saturation magnetization $M_S$ is predicted to be $M_S = gJ\ \mu_B = 2.0\ \mu_B$ for $SrRuO_3$. Measured values of $M_S$ are much less. Polycrystalline $SrRuO_3$ reaches about 0.85 $\mu_B$ [175, 178, 179] in low fields but continues to increase in higher magnetic fields. At 125 kOe it was noted that $M$ had reached 1.55 $\mu_B$ but had not yet saturated [178]. Early neutron diffraction derived a moment of 1.4 ± 0.4 $\mu_B$ [178] with no evidence for any antiferromagnetic order (spin canting). Recent theoretical investigations predict incomplete band splitting and a large moment of about 1.6 $\mu_B$ [176, 180]. Early explanations for the low value of $M_S$ included spin canting, band magnetism, and incomplete alignment of the magnetization due to magnetocrystalline anisotropy [178]. The present work shows that $SrRuO_3$ has a large saturation moment of 1.6 $\mu_B$ as predicted by these calculations.

Single crystals of $SrRuO_3$ can be grown from a $SrCl_2$ flux [36]. The resistivity of such crystals is consistent with the results on polycrystalline samples. Magnetization and magnetic torque measurements of single crystals [181-183] showed that $SrRuO_3$ has a high cubic anisotropy with $\langle 110 \rangle$ (cubic cell) being the easy direction, with nearly square hysteresis loops. Some previous measurements reported for single crystals are questionable, for example, the measured value of $M_S = 1.1\ \mu_B$. In this work new magnetization data is shown to clarify these points.

**Experimental**

Single crystals of $SrRuO_3$ were grown by slow cooling in a $SrCl_2$ flux [36, 184]. Polycrystalline $SrRuO_3$ was prepared from stoichiometric quantities of $SrCO_3$ and Ru metal repeatedly reacted at 1260°C, and was used as the source material for crystal growth. The $SrCl_2$ was dried in air at 110°C. A mixture with approximate weight ratio 1:20 of $SrRuO_3$:$SrCl_2$ was melted in a platinum crucible with lid at 1260°C for 94 hrs. The sample was cooled to 800°C at 1°/hr and then to room temperature at ~40°/hr. Crystals of $SrRuO_3$ less than 1 mm in diameter were found at the bottom of the crucible after removing the flux with water. Most crystals were cubo-octahedron shaped and grew with a 3-fold symmetric axis (presumably [111]) perpendicular to the Pt surface. The (cubic) crystal orientation was determined by the symmetry of the faces. The actual orthorhombic symmetry [177] was confirmed by powder x-ray diffraction and Transmission Electron Microscopy.

Magnetization in fields up to 70 kOe was measured using a Quantum Design $MPMSR_2$ SQUID magnetometer. The samples were attached to a plastic straw with a small amount of Apiezon N vacuum grease. Samples were oriented on the straw visually using crystal faces which had obvious 2-, 3- and 4-fold rotational symmetry. At room temperature, the crystal can rotate in a large field to align a paramagnetic easy direction with the field. This was used as a final adjustment when aligning along the easy $\langle 110 \rangle$ directions. At low temperatures, the grease solidifies so the sample cannot rotate. The accuracy of the magnetization measurement was estimated by measuring a sphere of yttrium iron garnet described [1]. For the measurements reported here, the sample holder was fixed at the rotation angle which gave the maximum magnetization. Several crystals were measured.

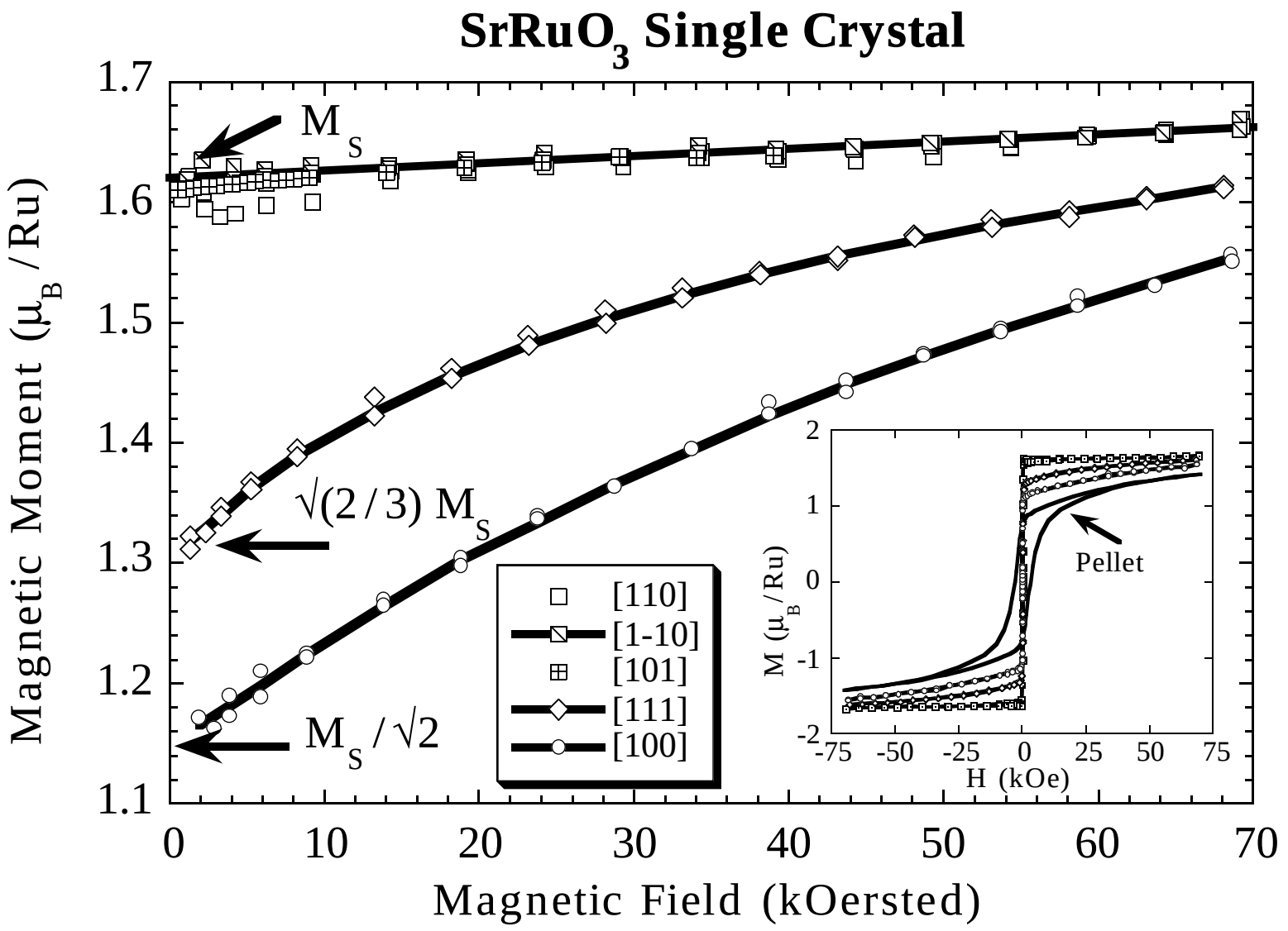

**Figure 1. Magnetization at 5 K of $SrRuO_3$ single crystal along several crystallographic directions showing strong cubic but not uniaxial magnetocrystalline anisotropy. Inset shows the full hysteresis loop of the single crystal data along with that of a polycrystalline pellet for comparison.**

## Results

Figure 1 shows hysteresis loops at 5 K for $SrRuO_3$ single crystal in various crystallographic directions and a polycrystalline pellet. The crystals have a low coercive field (≈ 10 Oe) compared to that of the polycrystalline pellet (3000 Oe). But more importantly, the crystals show very little hysteresis while the pellet displays noticeable hysteresis even in fields greater than 40 kOe.

The rapid, linear approach to saturation (with respect to the applied field) found in all directions can be attributed to demagnetization. Until the sample becomes fully magnetized, the demagnetization field $H_d$ is equal to the applied field $H_a$ resulting in an internal field of zero ($H_i = H_a - H_d$). From this slope ($M = H_a/4\pi N$), one can calculate the demagnetization factor $N$ and therefore calculate the internal field. The measured demagnetization factors $N$, ($H_d = 4\pi NM$) are 0.25, 0.28, 0.31, 0.49, 0.66 for the [110], [1-10], [101], [100] and [111] directions respectively. These seem reasonable considering the shape of the crystals. Since a uniaxial magnetocrystalline anisotropy will also give a linear increase in $M$ if $H$ is applied along the hard direction, it is difficult to distinguish it from the effect of the demagnetization field in this study. Since the demagnetization field can be as large as a thousand Oersted, one can only conclude that the uniaxial anisotropy field is less than a thousand Oersted, which is considerably less than that ( > 50 kOe) reported previously [182].

In the [110] direction $SrRuO_3$ rapidly approaches saturation and then remains relatively constant (square hysteresis loop), as is expected for a magnet with a magnetic field along the easy direction. The crystal was also measured in the [1-10] and [101] directions which would be equivalent by symmetry to the [110] if the crystals were cubic. Since the magnetization is the same along these $\langle 110 \rangle$ type directions, it can be concluded that the magnetic properties of single crystal $SrRuO_3$ are essentially cubic, *i.e.* only cubic magnetocrystalline anisotropy is detected. Beyond this initial saturation along the easy $\langle 110 \rangle$ directions, there is a small but measurable increase in the magnetization which is linear in magnetic field and has a slope of $6 \times 10^{-7}$ $\mu_B$/Oe.

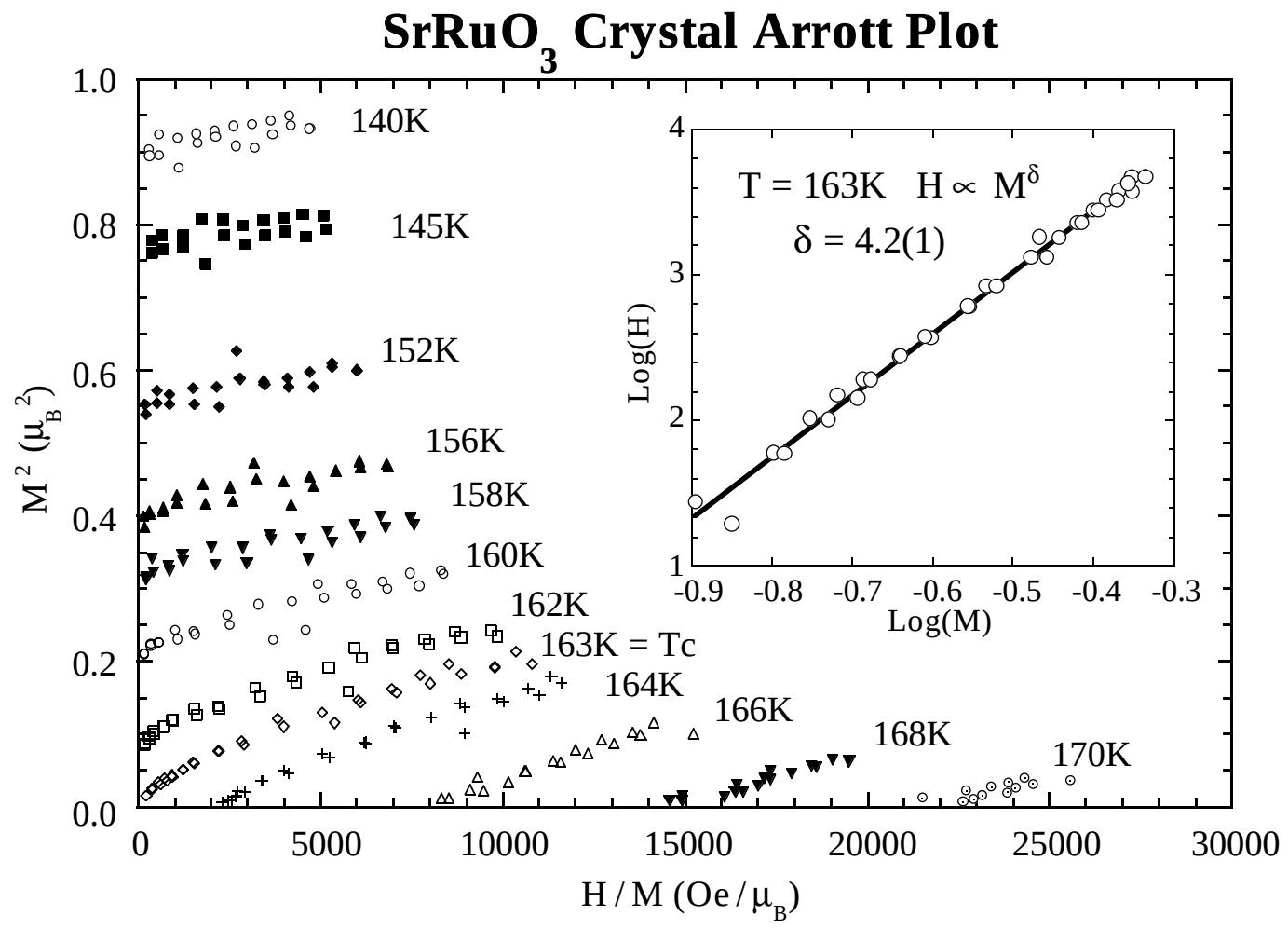

**Figure 2. Arrott Plot of $SrRuO_3$ single crystal along easy [110] direction. Inset, critical isotherm ($T$ = 163K ≈ $T_C$) on a log scale fit to $M^\delta \propto H$ with $\delta = 4.2$.**

The $H = 0$, $T = 0$ saturation moment $M_S$ found along the easy ⟨110⟩ directions are about 1.62 $\mu_B$/Ru. In the remnant state ($H$ reduced to zero), the magnetization should lie along the nearest easy direction. Thus the expected remnant ($H = 0$) magnetization along the [100] or [111] direction is simply the cosine of the angle it makes with the closest ⟨110⟩ direction. For ⟨100⟩ the expected remnant magnetization is $M_S/\sqrt{2}$, and for ⟨111⟩ directions $\sqrt{(2/3)}M_S$ is expected. These are extremely close to the experimental values (Figure 1).

In the other primary directions, there is a nonlinear approach to saturation which is characteristic of materials with cubic anisotropy. The magnetocrystalline anisotropy constants $K_1$ and $K_2$ can be estimated from these magnetization curves [185]. The magnetocrystalline anisotropy energy $E$ can be defined in terms of the cosine of the angle $\mathbf{M}$ makes with the three crystal axes: $\alpha_i = \mathbf{x}_i \bullet \mathbf{M}/M$. For a cubic material $E = K_1\,(\alpha_1^2\alpha_2^2 + \alpha_2^2\alpha_3^2 + \alpha_3^2\alpha_1^2)$ +

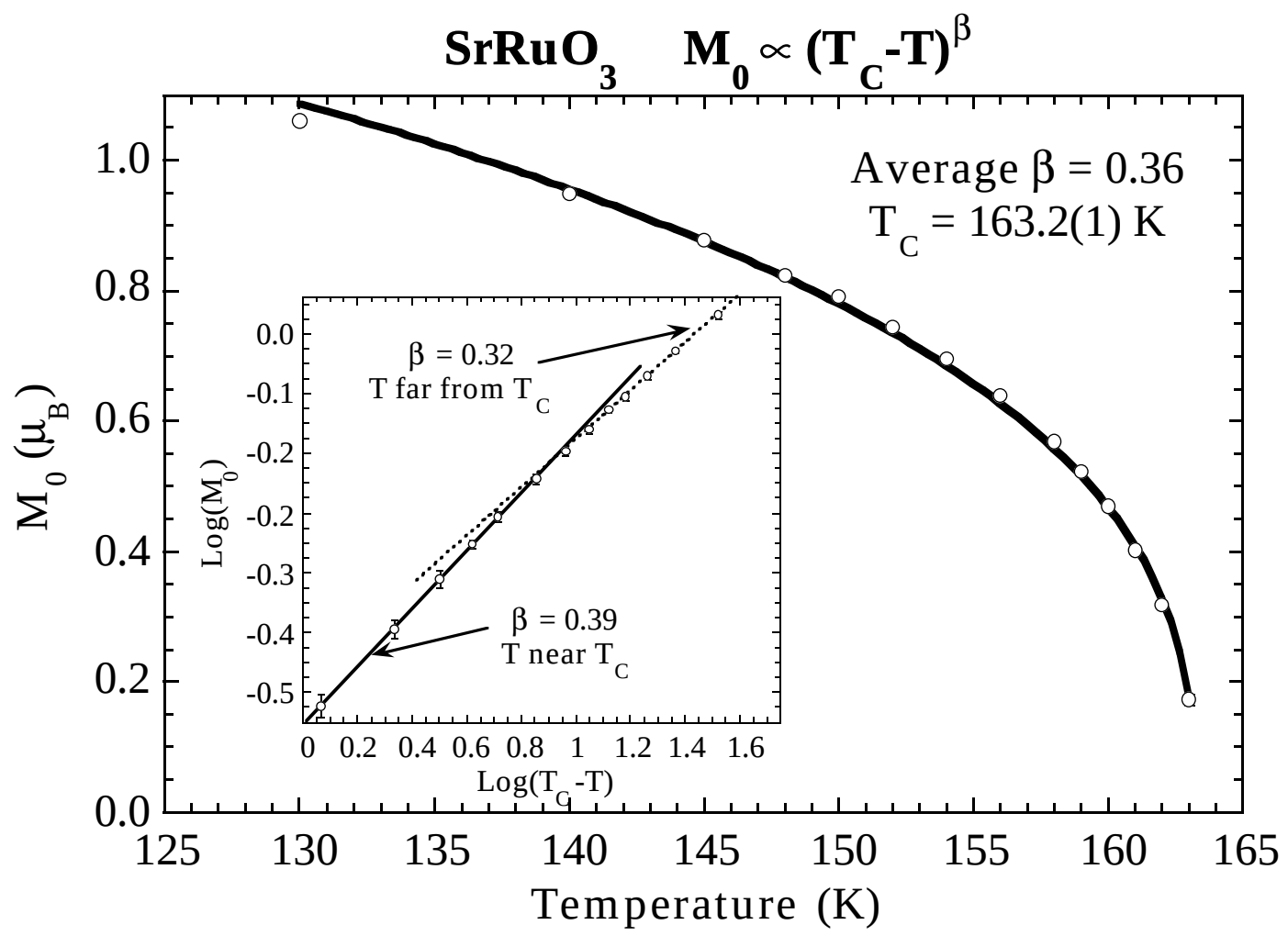

**Figure 3. Zero field magnetization $M_0$ of $SrRuO_3$ single crystal along easy [110] direction. Solid line shows the fit to $M_0(T) \propto (1 - T/T_C)^\beta$ with $\beta = 0.36$. Inset showing the same data on a log plot. The critical exponent $\beta$ appears to change from Heisenberg-like $\beta = 0.39$ near $T_C$ to Ising-like $\beta = 0.32$ as $T$ decreases.**

$K_2 (\alpha_1^2\alpha_2^2\alpha_3^2) + \ldots$ . In $SrRuO_3$ the magnetic easy axes are $\langle 110\rangle$, which requires $K_1 < 0$. Similarly, $\langle 100\rangle$ are the hard axes ($\langle 111\rangle$ are intermediate) which implies $2.25|K_1| < K_2 < 9|K_1|$. The [100] magnetization should intersect that of the easy axis [110] at $H_a = -2K_1/M_S$. An extrapolation of the [100] $M$ *vs.* $H$ curve gives $-2K_1/M_S \approx 109$ kOe. Alternatively, the area between the [100] and the [110] $M$ *vs.* $H$ curves should be equal to $-K_1/4$. This method gives a value for $-2K_1/M_S \approx 96$ kOe. The area method can also be used with the [111] curve to estimate $2K_2/M_S \approx 540$ kOe.

In the easy [110] direction, $M$ *vs.* $H$ curves were measured for various temperatures and are shown in Figure 2 as $M^2$ vs. $H/M$ (Arrott plot). The isotherms below $T_C$ should be approximately linear and intersect $H/M = 0$ at $M_0$.

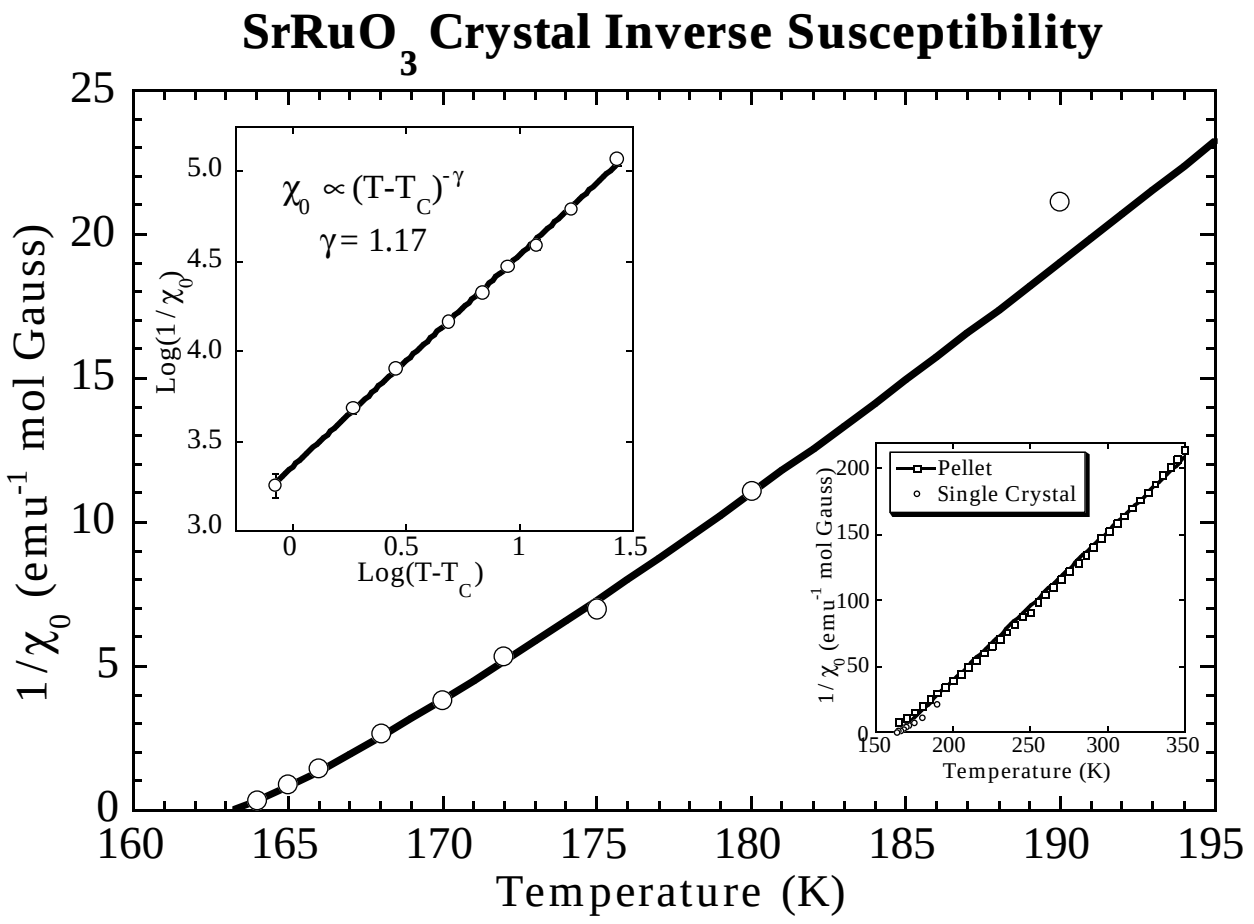

**Figure 4. Zero field inverse susceptibility $1/\chi_0$ of $SrRuO_3$ single crystal along easy [110] direction. Solid line shows the fit to $1/\chi_0(T) \propto (1 - T/T_C)^\gamma$ with $\gamma = 1.17$ and $T_C = 163.2$ K. The inset shows the same data on a log plot.**

From these $M_0(T)$, shown in Figure 3, the critical exponent $\beta \approx 0.36$ and $T_C$ = 163.2 ± 0.2 K can be estimated by fitting $M_0(T) \propto (1 - T/T_C)^\beta$ in the critical region. The fit with a single value for $\beta = 0.36$ is poor considering the apparent precision of the data. Closer to $T_C$, the data fit better with a larger $\beta \approx 0.39$ which is close to that predicted in the 3-*d* Heisenberg model ($\beta$ = 0.38). Farther from $T_C$, the exponent is smaller, $\beta \approx 0.32$, near that predicted by the Ising model ($\beta$ = 0.33). It is possible that this is due to a crossover from Heisenberg to Ising behavior at $T$ decreases from $T_C$. A similar model has been proposed to explain measurements of thin film $SrRuO_3$ [73].

The $T > T_C$ isotherms of Figure 2 should intersect $M^2 = 0$ at $1/\chi(T, H = 0) = 1/\chi_0$. The critical exponent $\gamma$ is then found from $1/\chi_0 \propto$ $(T/T_C - 1)^{\gamma}$. From these data (Figure 4) $\gamma = 1.17 \pm 0.02$ is estimated. The plot of $1/\chi$ vs. $T$ should be close to linear for high temperatures.

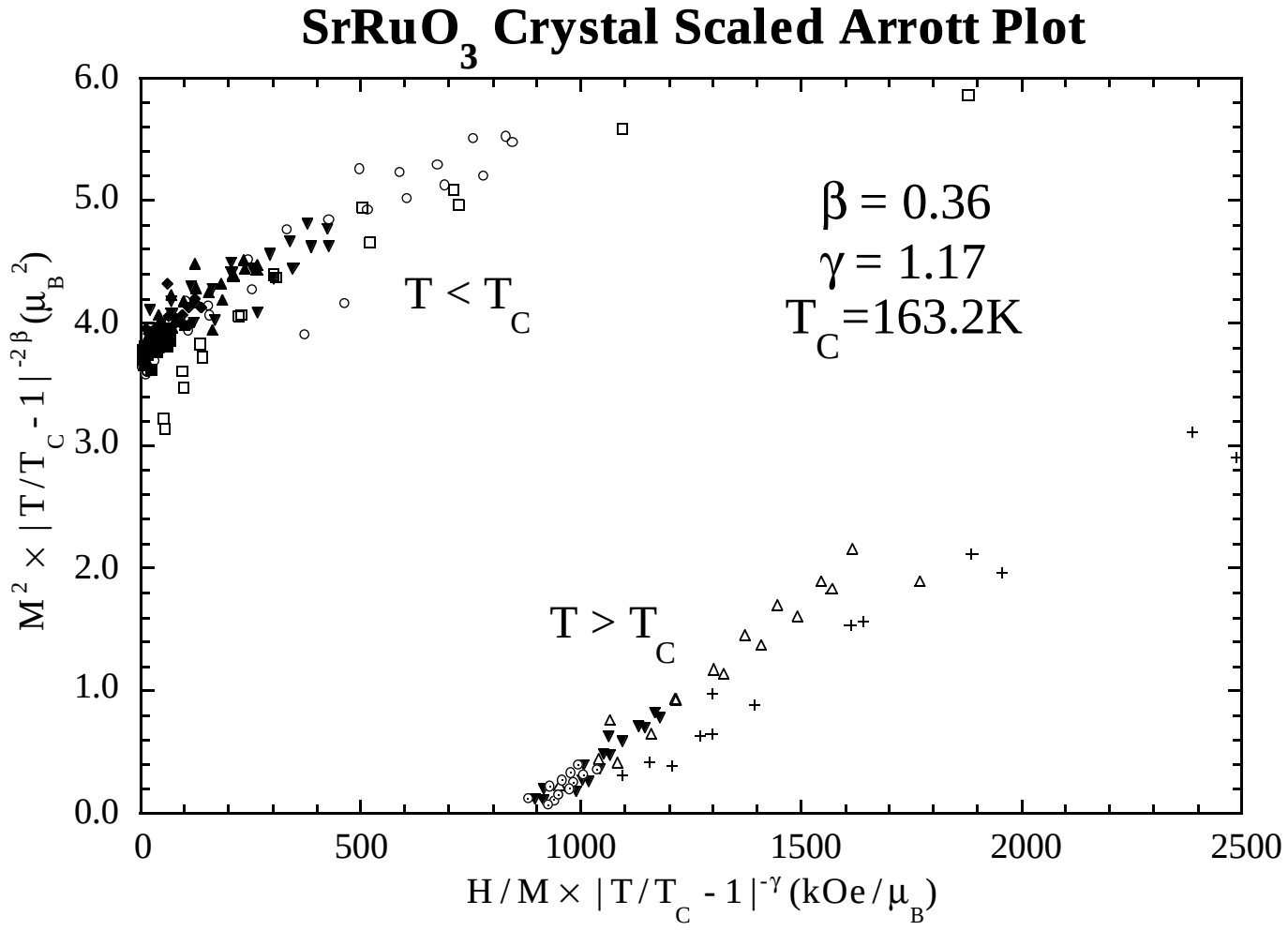

**Figure 5. Scaled Arrott Plot of $SrRuO_3$ single crystal along easy [110] direction with $\beta$ = 0.36 and $\gamma$ =1.17. Symbols are the same as those used in Figure 2.**

The $M(H, T)$ data can be replotted using the scaling hypothesis [90] with $\beta = 0.36$ and $\gamma = 1.17$. According to the hypothesis, the magnetic equation of state in the critical region depends only on the scaled variables $H/|T_C/T - 1|^{\beta+\gamma}$ and $M/|T_C/T - 1|^{\beta}$. A plot of the scaled $M^2$ and scaled $H/M$, shown in Figure 5, will then have only two curves: one branch for the $T < T_C$ data and another for $T > T_C$. Not all the curves fit on a single line, this is due to the apparent change in $\beta$ as discussed above.

The critical isotherm ($T = T_C$) should obey the relation $M^{\delta} \propto H$, where according to the scaling relation $\delta = \gamma/\beta + 1$. From the previously measured values of $\beta$ and $\gamma$, the exponent $\delta$ is therefore expected to be $4.3 \pm 0.5$. This is in good agreement with the measured value (Figure 2) $\delta = 4.2 \pm 0.2$.

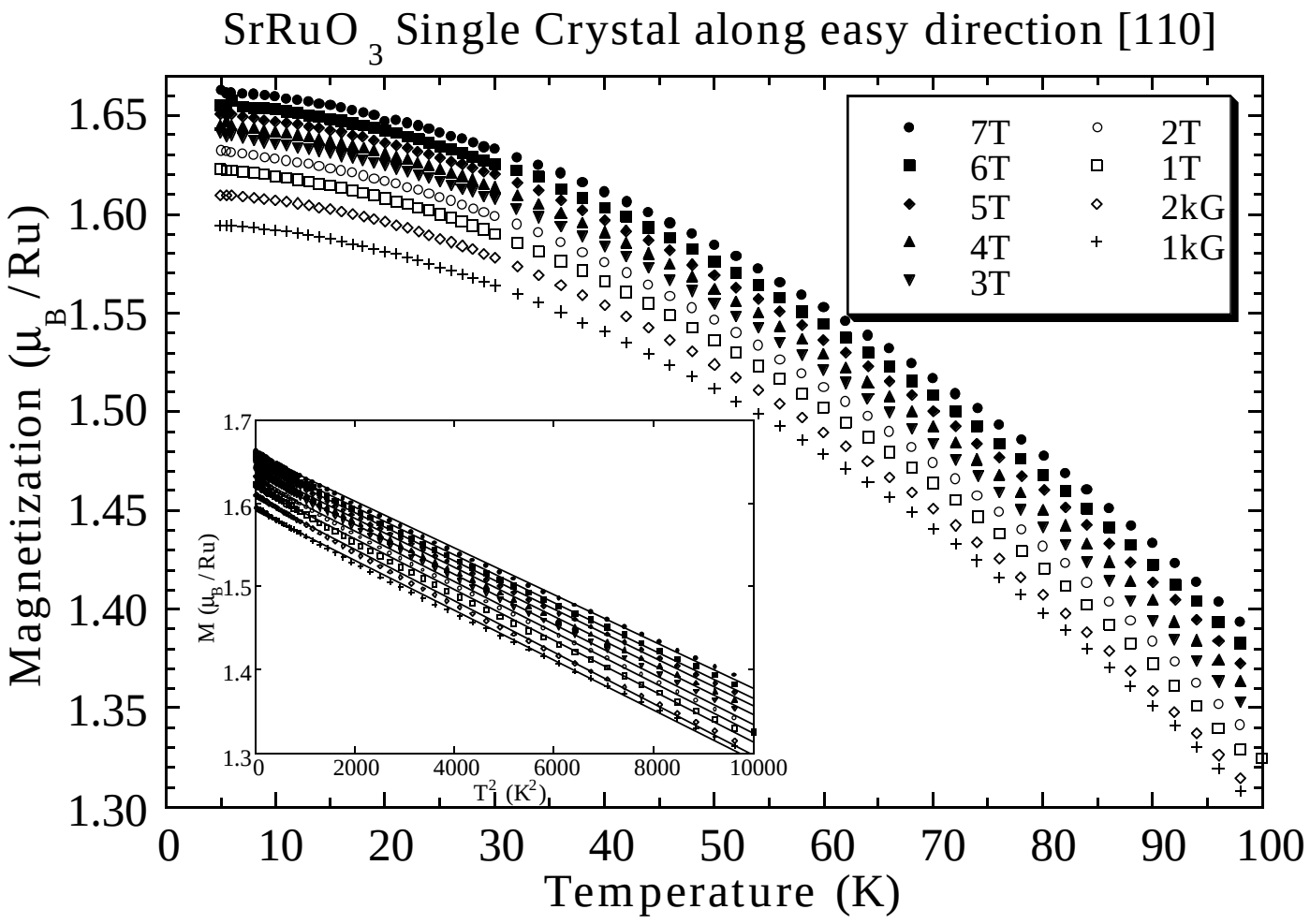

**Figure 6. Magnetization as a function of temperature of $SrRuO_3$ single crystal along easy [110] direction. Inset shows the approximate $T^2$ dependence of the magnetization.**

The temperature dependence of $M$ along the easy direction is shown in Figure 6 for various applied fields. The demagnetization field of $4\pi NM \approx 800$ Oe should be subtracted from the applied field to give the internal field. Between 5 K and 100 K, and for internal fields from 0.2 kOe to 68 kOe the magnetization is well approximated by $M = M_0(1 - (T/\Theta_2)^2)$ where $M_0$ and $\Theta_2$ are fitting parameters. Other possible analyses are discussed below.

## Discussion

The low temperature magnetization data (Figure 1) of bulk $SrRuO_3$ can be explained with $\langle 110 \rangle$ easy directions and a large cubic, but very little uniaxial, magnetocrystalline anisotropy. The low coercivity of the crystals (~10 Oe), compared to the 3 kOe coercivity found in polycrystalline samples, provides highly reversible and square hysteresis curves. There is no indication of a magnetic multi-domain structure [183].

The cubic magnetocrystalline anisotropy in $SrRuO_3$ is very large, as reported previously [181, 183]. Typical values of $K_1$ for cubic 3-$d$ ferromagnets are 100 times smaller than that found for $SrRuO_3$. Such large values of $K_1$ usually refer to uniaxial anisotropy, for instance in hexagonal materials, which is a lower order effect ($K_1$ refers to the first non-zero anisotropy constant). This large anisotropy probably results from the strong spin-orbit coupling of the heavy Ru atom, which also gives $SrRuO_3$ a strong Kerr effect [186].

From measurements along different $\langle 110 \rangle$ directions, no indication of uniaxial magnetocrystalline anisotropy [182] is found, although it is allowed from the orthorhombic symmetry. This might be expected since the crystallographic unit cell lengths [177] vary by only 0.03%, and the angles by 0.4% from the perfect cubic ones. The distortion from cubic is primarily due to a rotation of the $RuO_6$ octahedra, which alters the symmetry much more than the shape of the unit cell [177]. In the crystals reported here, the orthorhombic cell was confirmed using TEM. Because the unit cell is only slightly distorted, the few reports claiming cubic [184] or tetragonal [182] crystallographic symmetry without supporting evidence, should be reevaluated in this context. Significant uniaxial anisotropy in thin films of $SrRuO_3$ may result from growth induced anisotropy, as was found in films of magnetic garnets used in magnetic bubble technology [188].

Due to the large magnetocrystalline anisotropy, the saturation moment of $SrRuO_3$ is difficult to measure. In directions other than $\langle 110 \rangle$ the magnetic moment does not saturate even in fields of several 10 kOe. The strain and small particle size of a polycrystalline sample apparently makes it even more difficult to saturate than the hard direction in a single crystal (Figure 1). Polycrystalline $SrRuO_3$ has a remnant magnetization ($H$ = 0) of $M_S \approx 0.85\ \mu_B$, which increases non-linearly past 1.55 $\mu_B$ ($H$ = 125 kOe) [178]. Clearly a magnetically soft single crystal with $H$ along the easy direction is needed to measure $M_S$. In a previous experiment [181] on single crystal $SrRuO_3$ with a square hysteresis loop, $M_S$ = 1.1 $\mu_B$ at $H$ = 0 was reported. However, that value of $M_S$ is clearly too small since by 17 kOe it is smaller than the value for a polycrystalline sample [181]. Our measured value of $M_S$ = 1.6 $\mu_B$ ($H$ = 0) is consistent with the high field polycrystalline results.

The magnetic critical exponents measured here are in the range typically seen in large moment ferromagnets and expected theoretically for 3-dimensional ferromagnets. Experimental values for the critical exponent β in Fe, Ni and YIG [92] are 0.37 ± 0.02, which are near the theoretical values (Ising β = 0.33, Heisenberg β = 0.36). The related metallic ferromagnets $La_{0.5}Sr_{0.5}CoO_3$ [92] has β = 0.361. The apparent decrease in β as $T$ decreases from $T_C$, may be due to the large magnetocrystalline anisotropy. A similar effect has been seen in thin film $SrRuO_3$, where it is suggested that the magnetocrystalline anisotropy induces a crossover from Heisenberg to Ising behavior. The weak, itinerant electron ferromagnet $ZrZn_2$ has mean field critical exponents β = 0.5 [93]. The most prominent theory on itinerant electron ferromagnetism by Moriya [84, 85] predicts a $T_C^{4/3} - T^{4/3}$ dependence on the magnetization, which is essentially β = 1.

The critical exponents γ and δ are also typical for large moment ferromagnets (Fe, Ni and YIG [92] γ = 1.2 ± 0.2) as opposed to those found for the weak, itinerant electron ferromagnet $ZrZn_2$ which has mean field critical exponents γ = 1.0 and δ = 3 [93]. The three dimensional Ising and Heisenberg models predict γ = 1.24, δ = 4.8 and γ = 1.39, δ = 4.8 respectively. The critical exponents measured here also obey the scaling relation $\delta = 1 + \gamma/\beta$.

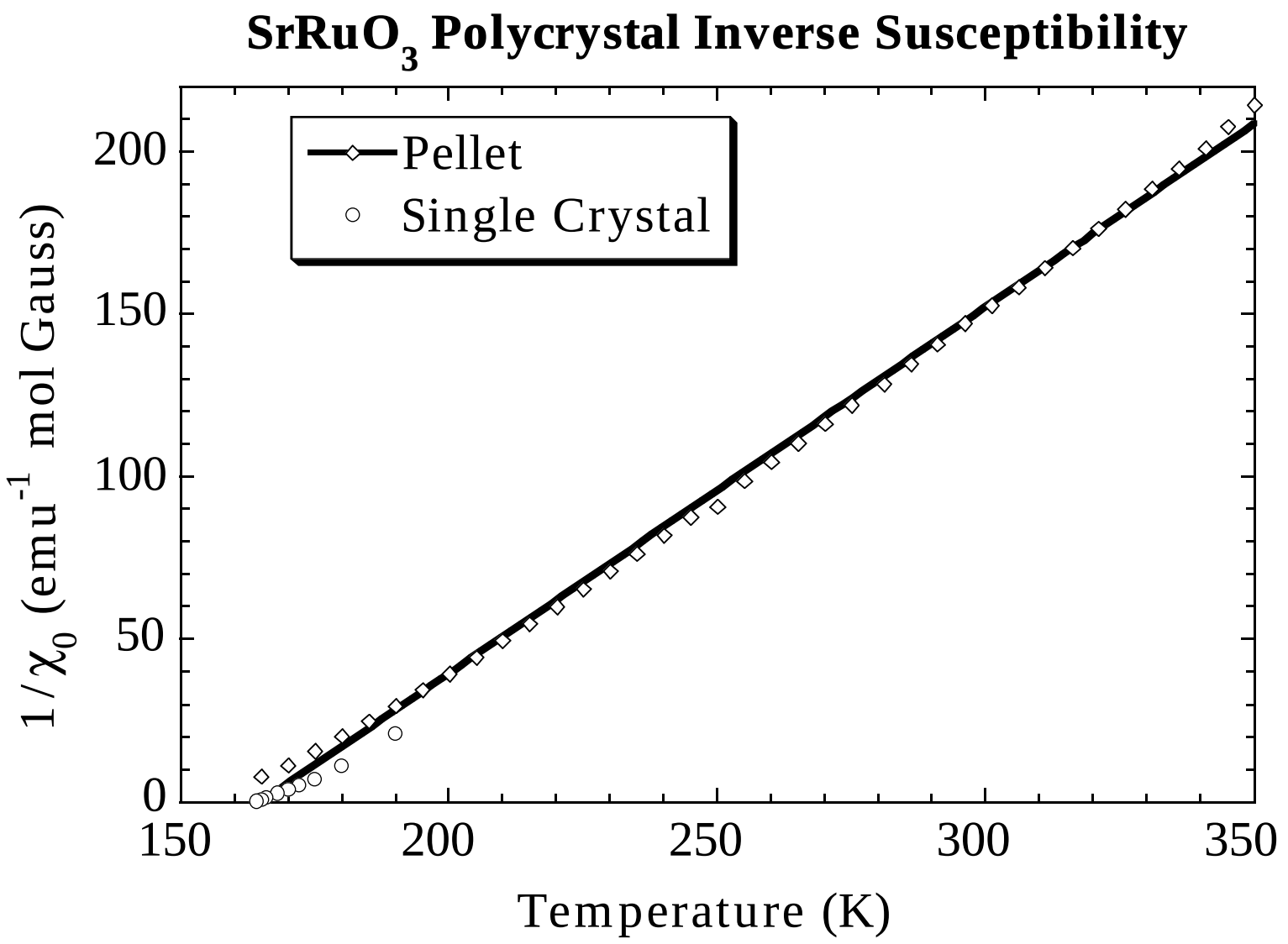

**Figure 7. Inverse magnetic susceptibility ($1/\chi = M/H$) at $H$ = 10 kOe of polycrystalline $SrRuO_3$ compared to the single crystal data from Figure A- 4. The solid line is the straight-line fit with $T_C$ = 165K which demonstrates the slightly positive curvature of the data.**

A positive curvature persists in the plot of $1/\chi$ *vs.* $T$ (Figure 7) even at higher temperature where it should be linear for a Curie-Weiss ferromagnet. Such a curvature can be caused by a temperature independent term in the susceptibility $\chi = \chi_{\mathrm{Const}} + C/(T - \Theta)$, where $C/(T - \Theta)$ is the Curie-Weiss susceptibility and is always positive ($T > \Theta$). A $\chi_{\mathrm{Const}} < 0$ of about $-4 \times 10^{-4}$ emu $G^{-1}$ $mol^{-1}$ will provide the observed curvature, and has been independently observed elsewhere [175]. The temperature independent term $\chi_{\mathrm{Const}}$, should contain a positive contribution due to Pauli paramagnetism. This can be estimated from measurements of the linear term of the specific heat [176], giving $\chi_{\mathrm{Pauli}} \approx +4 \times 10^{-4}$ emu $G^{-1}$ $mol^{-1}$. The Landau diamagnetism should be negative and for simple band structures is smaller than $\chi_{Pauli}$ (for free electrons $\chi_{Landau} = -\chi_{Pauli}/3$). The core diamagnetism can be estimated from tables of experimental values [76] giving $\chi_{Core} = -0.7 \times 10^{-4}$ emu $G^{-1}$ $mol^{-1}$. The sum of these theoretical estimates $\chi_{Const} = \chi_{Pauli} + \chi_{Landau} + \chi_{Core}$ is, however, still positive while the experimental value appears to be negative. In the critical region, $1/\chi \propto (T/T_C - 1)^{\gamma}$ with $\gamma = 1.17$, provides a positive curvature in the plot of $1/\chi$ *vs.* $T$. Since the mean field exponent $\gamma = 1$ is expected to be valid far from $T_C$, some other mechanism must provide the effective $\gamma > 1$ observed at these higher temperatures.

Magnetic excitations which become thermally induced as the temperature is raised above $T = 0$ reduce the magnetization from the ground state value. The exponential decrease predicted by the mean field model has some qualitative value but is never in good agreement with experiment. Collective spin wave excitations and single particle (Stoner) excitations both decrease the magnetization according to a power law $M = M_S(1 - (T/\Theta_n)^n)$ which is in accord with experiments where $n \approx 2 \pm 1$ and $\Theta_n$ is of the order $T_C$.

The limiting low $T$, $H = 0$ behavior of collective, spin wave excitations (section 3.2.2.2.8) predicts $n = 3/2$ and for $SrRuO_3$, $\Theta_{3/2} \approx 2.42 T_C = 400$ K. The Stoner theory of single ($k$-space) particle excitations (section 3.2.2.2.2 of [1]) predicts $n = 2$ and $\Theta_2 \approx 1.41\ T_C$ (= 230 K for $SrRuO_3$). As the temperature and magnetic field is raised, these models require further corrections. For example, higher order corrections such as an additional $n = 5/2$ term in predicted in the spin wave theory. For $H \neq 0$, the low temperature spin-wave excitations become quenched which can have the effect of increasing the average value on $n$ [97]. The generalized model of spin fluctuations in ferromagnets ([84] section 3.2.2.2.3 in [1]) includes both types of interacting magnetic excitations and predicts $n = 3/2$ for $H = 0$, $T = 0$ but also $n \approx 2$ in calculations [86].

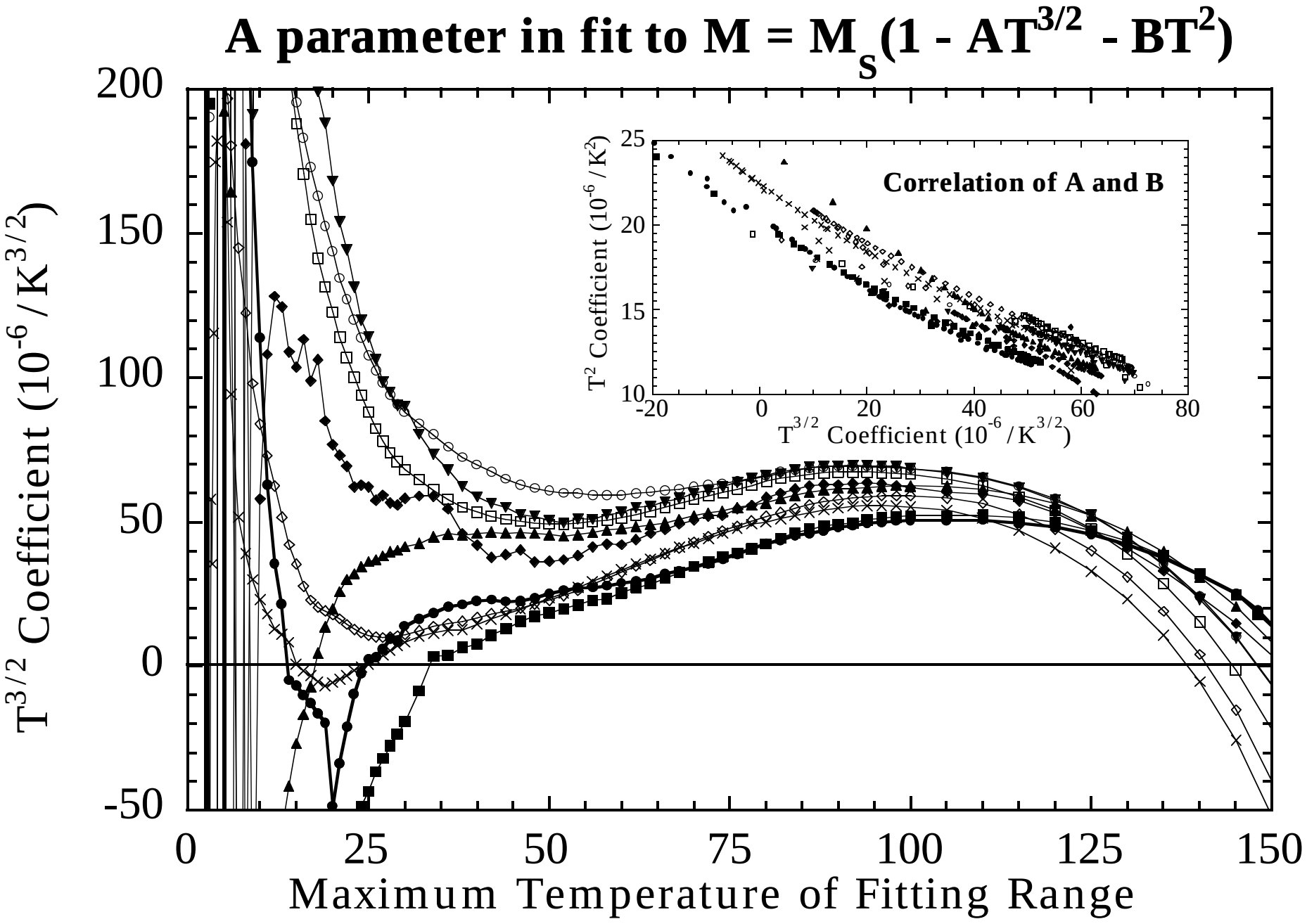

**Figure 8. Variation of the $T^{3/2}$ parameter in fitting the magnetization data of single crystal $SrRuO_3$ to $M = M_S\,(1 - AT^{3/2} - BT^2)$ as the fitting range is increased. The upper inset shows the correlation of the $A$ and $B$ parameters. In the region where $A$ is relatively stable (around $T_{max}$ = 60 K), $A$ decreases as $T_{max}$ is lowered. The symbols are the same as those used in Figure 9.**

Experimental results on metallic ferromagnets with substantial saturation moments such as Fe, Ni [88, 189-192] and $La_{0.67}Sr_{0.33}MnO_3$ ([97], section 4.1.1 in [1]) can display $n = 3/2$ consistent with the spin wave stiffness determined by neutron diffraction if significant corrections are included and only low temperature data is analyzed. In contrast “weak” itinerant-electron ferromagnets such as $ZrZn_2$, $Ni_3Al$ and $Sc_3In$, show $n = 2$ [86, 193-195] over a wide temperature range.

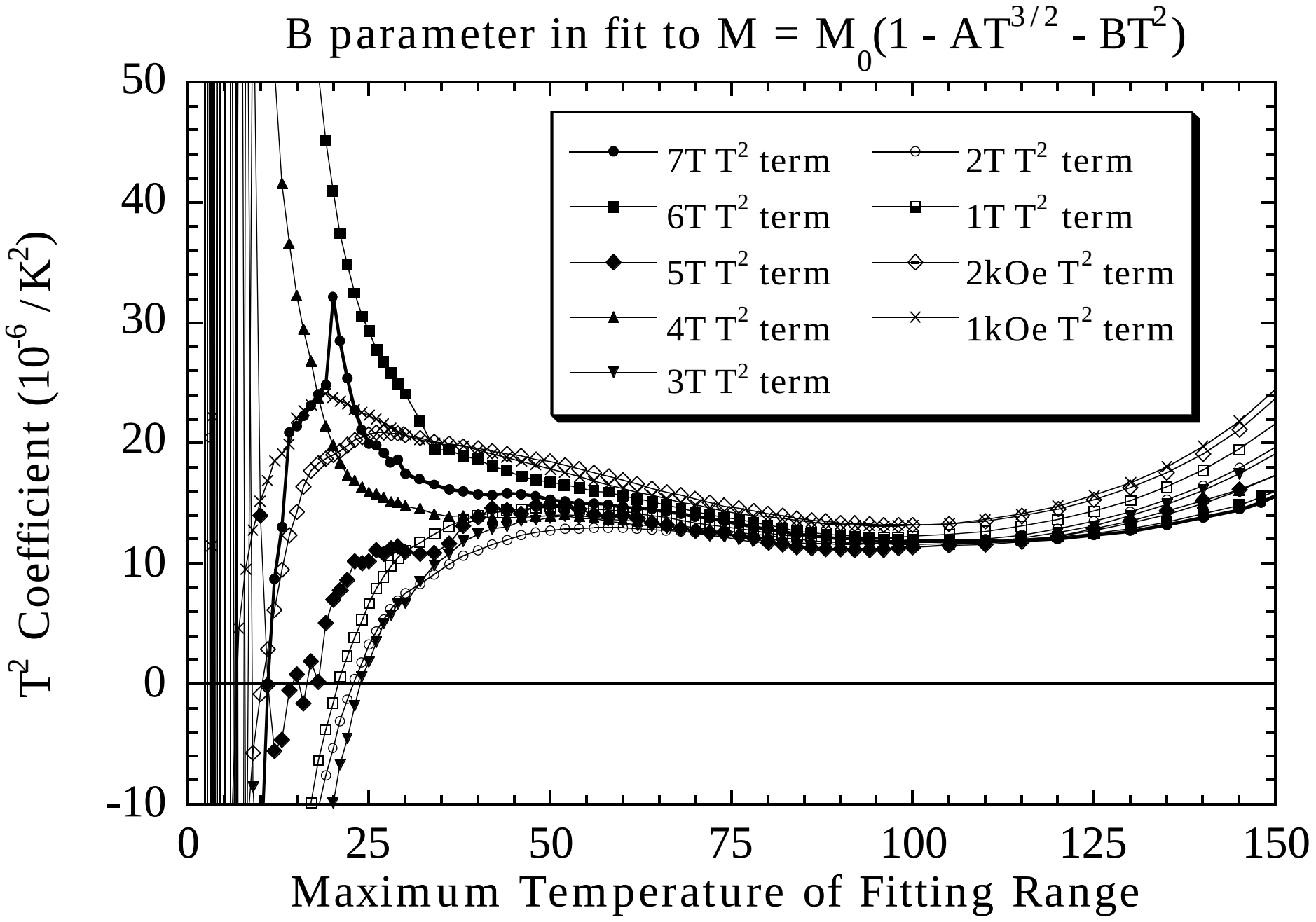

**Figure 9. Variation of the $T^2$ parameter in fitting the magnetization data of single crystal $SrRuO_3$ to $M = M_S\ (1 - AT^{3/2} - BT^2)$ as the fitting range is increased. In the region where $B$ is relatively stable (around $T_{max}$ = 60 K), $B$ increases as $T_{max}$ is lowered.**

As shown above (Figure 6) the magnetization of $SrRuO_3$ single crystal can be well described by $M = M_S(1 - (T/\Theta_n)^n)$ with $n = 2$ in the temperature range of reliable measurement, 5 K to 100 K. A fit with $n = 3/2$ in this temperature range is unsatisfactory.

If one assumes $n = 2$ arises from Stoner excitations and collective excitations result in $n = 3/2$, then one may expect the variation of magnetization with temperature to fit with a combination of the $n$ = 2 and $n$ = 3/2 terms, $M = M_S\,(1 - AT^{3/2} - BT^2)$ [150] at low temperatures. Since an additional parameter is included, an improvement of the fit does not necessarily prove the significance of the added parameter. Furthermore, the functions $T^{3/2}$ and $T^2$ are very similar making the fitting parameters highly correlated.

The magnetization data for single crystal $SrRuO_3$ were fit to a polynomial expression $M = M_S\,(1 - AT^{3/2} - BT^2)$ from $T$ = 2 K to $T = T_{max}$. As expected, the $A$ and $B$ parameters are highly correlated (Figure A- 8) with $\Delta B \approx 2.2 \times 10^{-5}$ K$^{-2}$ - 0.2 K$^{-1/2}$ $\Delta A$. The parameters $A$ and $B$ can then be plotted as a function of $T_{max}$ (Figure 8 and Figure 9). For very low $T_{max}$ (< 30 K) the fit is unstable since the difference in the magnetization at $T < T_{max}$ becomes comparable to precision of the measurement. Thus, the divergence of the fitting parameters at low $T_{max}$ is an artifact, and not physical. At high temperatures, nearing the critical temperature, the polynomial expression is not expected to be valid, so the divergence at high $T_{max}$ can also be ignored. The flat, linear region centered around $T_{max}$ = 60K, is presumably the region where the fitting parameters may have physical meaning. From this region, $A$ and $B$ parameters can be extracted and extrapolated to the $T_{max}$ = 0 K values for comparison with the theory. This analysis was done for fitting the magnetization data both to $M = M_S\,(1 - AT^{3/2} - BT^2)$ and $M = M_S\,(1 - BT^2)$.

Implicit in the use of $M = M_S\,(1 - AT^{3/2} - BT^2)$ is the assumption that these two terms are theoretical low temperature expansions, *i.e.* they should become more accurate as $T$ approaches 0 K. As the temperature range for which the data is fit to this type of expression is decreased to lower temperatures, the fitting parameters should be stable, or only be slightly varying. In principle, the true low temperature form should gradually become more dominant, at the expense of the other terms, as the fitting range is decreased to lower temperatures.

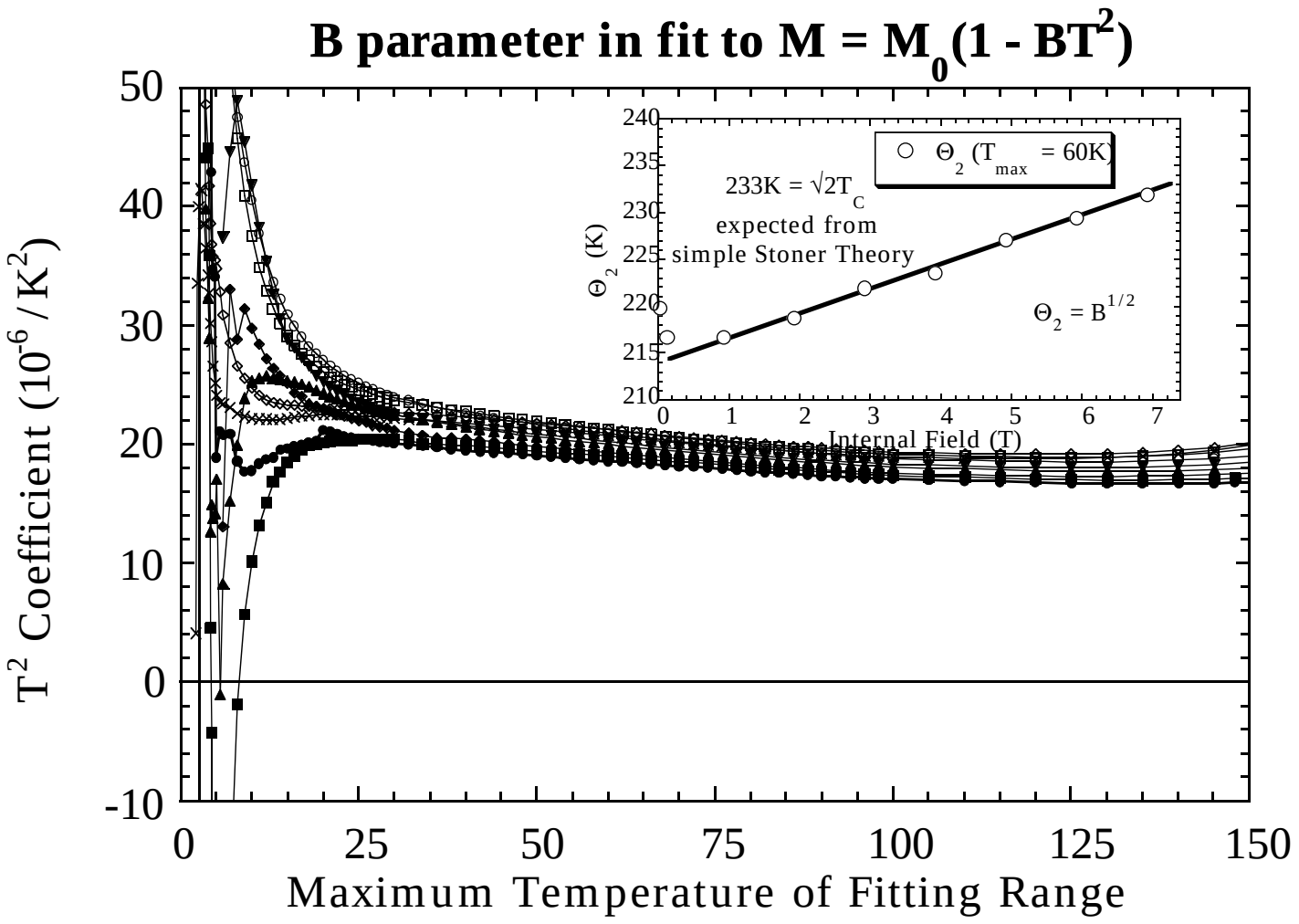

**Figure 10. Variation of the $T^2$ parameter in fitting the magnetization data of single crystal $SrRuO_3$ to $M = M_S\,(1 - BT^2)$ as the fitting range is increased. The parameter $B$ for this fit is more stable and constant than that shown in Figure 8. Inset, variation of $\Theta_2$ in a magnetic field.**

In fitting to the data on $SrRuO_3$ single crystal the $B$ parameter (Figure 9) is relatively stable and increases as lower temperature fitting ranges ($T_{max}$) are used. The $A$ parameter (Figure 8) is not as stable and appears to decrease as $T_{max}$ is lowered. If this trend were to continue, then the $T^{3/2}$ contribution would be very small at low temperatures. The removal of the $AT^{3/2}$ term further stabilizes the $B$ parameter (Figure 10), and makes $B$ more independent of $T_{max}$ than with $AT^{3/2}$ included. Thus, from this analysis of the magnetization data, there is little evidence for a large $AT^{3/2}$ contribution to the magnetization. This is in contrast with measurements on thin film [186, 187] which show $T^{3/2}$ dominating over $T^2$.

The two corrections to the spin wave $T^{3/2}$ theory mentioned above (finite magnetic field and higher order $T^{(2n+1)/2}$ terms) may not adequately explain the data presented here. The applied magnetic field will suppress spin wave excitations for $T < g\mu_B H / k_B$, resulting in an effective $n > 3/2$. If this were a significant effect, the $T^{3/2}$ fitting parameter $A$ should decrease as $H$ is increased particularly as $T_{max}$ approaches 0 K. Such a systematic decrease in $A$ is not obviously apparent (Figure 8). One cannot however, exclude the demagnetization or anisotropy field which are present even when the applied $H = 0$.

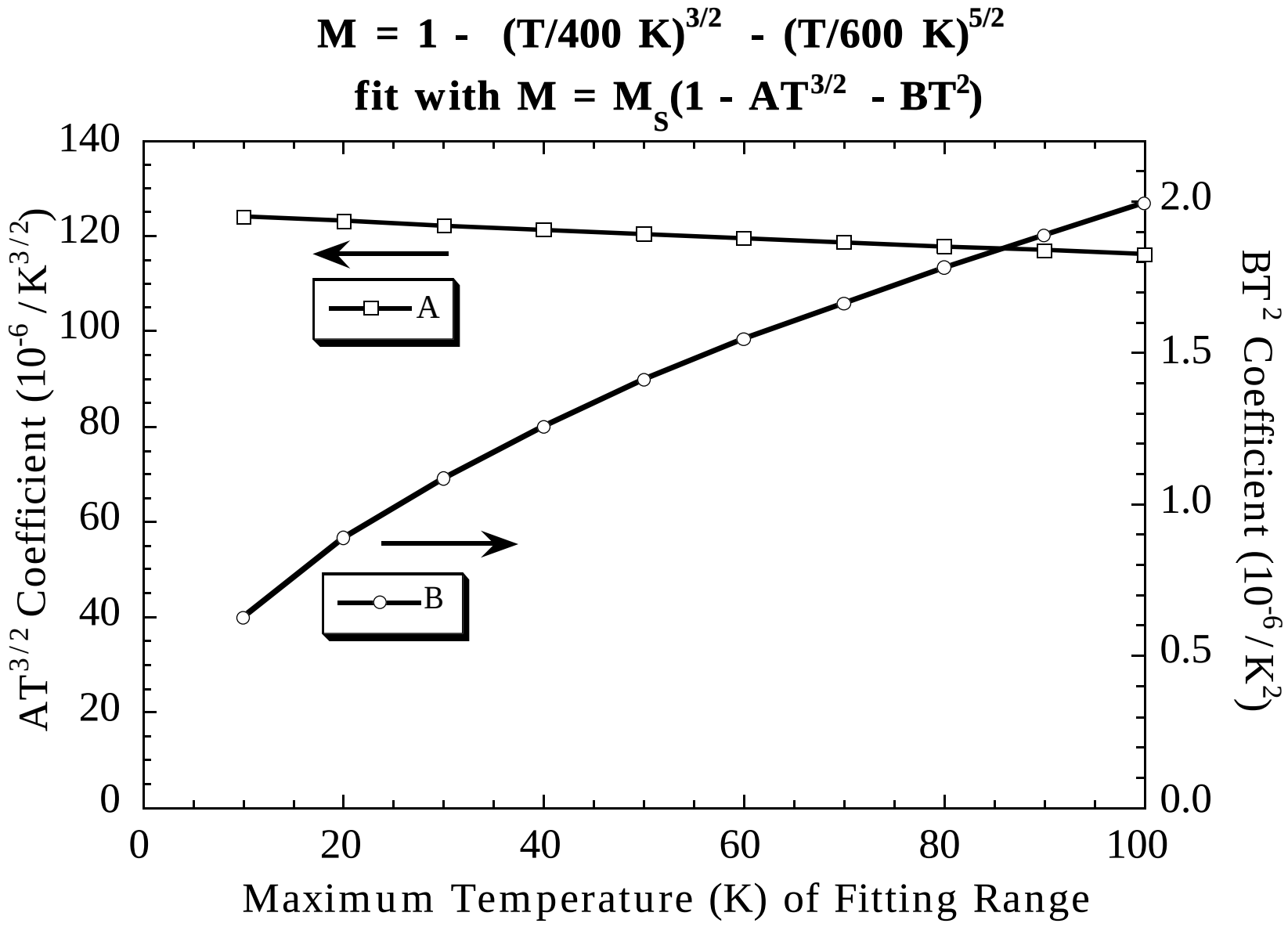

**Figure 11. Variation of *A* and *B* fitting parameters in the hypothetical case where the true magnetization is given by $T^{3/2}$ and $T^{5/2}$ terms.**

If the true magnetization were a sum of $T^{3/2}$ and $T^{5/2}$ terms, $M/M_S = (1 - (T/400\text{ K})^{3/2} - (T/600\text{ K})^{5/2})$ for instance, the data would fit well to $M/M_S = (1 - AT^{3/2} - BT^2)$. The fitting parameters $A$ and $B$ as a function of $T_{max}$ for this example are shown in Figure 11. The $A$ parameter is relatively constant, increasing slightly as $T_{max}$ approaches 0 K. In the limit $T_{max}$ approaches 0 K, $A = 125 \times 10^{-6}\text{ K}^{-3/2}$ or $\Theta_{3/2} = 400.3$ K, which is very close to the true value of 400 K. The $B$ fitting parameter is not nearly as stable and decreases as $T_{max}$ approaches 0 K. Since the exponent of the $BT^2$ fit (2) is smaller than 5/2, which is the exponent used in the example, a larger $B$ is needed to fit the data at higher temperatures. Thus, a dramatically decreasing $B$ as $T_{max}$ approaches 0 K, is indicative of a fitting exponent (2 in this example) smaller than the true exponent (5/2 in this example).

The data on $SrRuO_3$ shows the opposite effect: $B$ is relatively constant while $A$ decreases as $T_{max}$ approaches 0 K. This decrease of $A$ indicates that the exponent 3/2 is smaller than the true exponent, and therefore, the addition of higher order terms to the spin wave form of the magnetization will not fully explain the data on $SrRuO_3$.

Positive $B$ (and $A$) parameters can be converted to the $\Theta_n$ values used above via $\Theta_2 = B^{-1/2}$ and $\Theta_{3/2} = A^{-2/3}$. The $\Theta_2$ values extracted from fits described above are shown as a function of the internal magnetic field in Figure A- 10. The values of $\Theta_2(H)$ extrapolated to $T_{max} = 0$ are 20 K less than the corresponding $\Theta_2(H)$ for $T_{max}$ = 60 K. The single crystal $\Theta_2$ parameters increase only slightly as the field is increased, about 2.6 K/Tesla. These values for $\Theta_2$ are consistent with measurements of the field dependent heat capacity of polycrystalline $SrRuO_3$ [1] via the Maxwell relation

$$\left(\frac{\partial C_H}{\partial H}\right)_T = T\left(\frac{\partial^2 M}{\partial T^2}\right)_H.$$

## Conclusion

Various magnetic properties of single crystal $SrRuO_3$ along the magnetic easy direction have been measured. The saturation moment, $1.61\mu_B$/Ru, is larger than that reported previously for single crystal material but in accord with experiments on polycrystalline material and theoretical calculations. The crystals show extremely large cubic magnetocrystalline anisotropy while no substantial uniaxial magnetocrystalline anisotropy. The critical exponent $\beta$ changes from Heisenberg like 0.39 near $T_C$ to Ising like 0.32 further from $T_C$. The susceptibility exponent $\gamma$ = 1.17 persists to 2 $T_C$, where it is expected to decrease to 1. The magnetization from 5 K to 100 K is well described by $M = M_S(1 - (T/\Theta_n)^n)$ where $n \approx 2$. The addition of a $T^{3/2}$ term is not obviously significant.

## Acknowledgement

This work was performed in the KGB lab, Applied Physics, Stanford University 1993-1997. I thank Ted Geballe, Lior Klein, and Kathryn Moler for useful discussions. The work at Stanford was supported in part by the Air Force Office of Scientific Research and the Stanford Center for Materials Research under the NSF-MRL program. G.J.S. would like to acknowledge the generous support of the Hertz foundation. This manuscript has not been updated for publication since June 1997.